\documentclass[11pt]{article}
\usepackage{amsmath,amsthm}

\newcommand{\R}{{\mathbf R}} \newcommand{\N}{{\mathbf N}}
 \newcommand{\Z}{{\mathbf Z}}

\newcommand{\wt}{\widetilde }

\renewcommand{\epsilon}{\varepsilon } 
\newcommand{\g}{\gamma } 
\renewcommand{\rho}{\varrho } 
 
\renewcommand{\phi}{\varphi }
\renewcommand{\a}{\alpha }
\renewcommand{\b}{\beta }

\newcommand{\q}{{\rm q }}

\newtheorem{theorem}{Theorem}

\newtheorem{proposition}{Proposition}

\begin {document}
 \title{Quantum Approximation II. Sobolev Embeddings}

\author {Stefan Heinrich\\
Fachbereich Informatik\\
Universit\"at Kaiserslautern\\
D-67653 Kaiserslautern, Germany\\
e-mail: heinrich@informatik.uni-kl.de\\
homepage: http://www.uni-kl.de/AG-Heinrich}   
\date{}
\maketitle

\date{}
\maketitle
\begin{abstract}
A basic problem of approximation theory, the
approximation of functions from the Sobolev space $W_p^r([0,1]^d)$ 
in the norm of $L_q([0,1]^d)$, is considered from the point of view of quantum computation. 
We determine the quantum query complexity of this problem (up to logarithmic factors).
It turns out that  in certain regions 
of the domain of parameters $p,q,r,d$ quantum computation can reach a speedup of roughly 
squaring the rate of convergence of
classical deterministic or randomized approximation methods. There are other regions were 
the best possible rates coincide for all three settings.

\end{abstract}
\section{Introduction}
We are concerned with the study of numerical problems of 
analysis in the quantum model of computation. 
A series of papers dealt with scalar valued problems, that is, with problems,
whose solution is a single number. In \cite{NSW02} for the first time vector 
(function) valued problems were considered. The first analysis for such a type
of problem with matching upper and lower bounds was carried out in \cite{Hei03a}.

The present paper is a continuation of \cite{Hei03a}. We study one of the basic 
problems of
approximation theory, the approximation of functions from the Sobolev class 
$W_p^r([0,1]^d)$ in the norm of $L_q([0,1]^d)$, a problem which has received much 
attention
in the past in the classical settings (see the survey \cite{Hei93} and 
references therein).

Our results show that for $p<q$, the quantum model of computation can bring a speedup 
roughly 
up to a squaring of the rate in the classical (deterministic or randomized) setting.
On the other hand, for $p\ge q$, the optimal rate is the same for all three settings,
so in these cases there is no speedup of the rate by quantum computation.

Our method of analyzing the function approximation problem is similar to the one 
developed in \cite{Hei01b}, namely, we discretize the Sobolev embedding problem and show 
that a sufficiently precise knowledge about the discrete building blocks, the embeddings
of finite dimensional $L_p^N$  into $L_q^N$ spaces, leads to a full understanding of the
infinite dimensional problem. Although in a completely different setting, this is 
close in spirit to Maiorov's discretization technique from approximation theory 
\cite{Mai75}.
These finite dimensional embeddings were studied in \cite{Hei03a}, the results 
of which will be exploited here.
In this sense the present paper is related to  
\cite{Hei03a} in a similar way, as a previous paper \cite{Hei01b} 
on quantum integration in Sobolev spaces
was related to results on summation \cite{Hei01, HN01b}.

For an introduction and notation concerning the quantum 
setting of information-based complexity we refer to Section 2 of
\cite{Hei03a}. Some general results which we will use can be found in Section 3
of that paper. Finally, we also refer to \cite{Hei03a} for comments on the
bibliography.
In Section 2 of the present paper, which contains the main result, 
we study approximation of the embeddings of Sobolev classes $W_p^r([0,1]^d)$
into $L_q([0,1]^d)$. In Section 3 we shortly discuss the cost of the algorithm
in the bit model and compare the results to the classical settings.

\section{Approximation of Sobolev Embeddings}
Let $D=[0,1]^d$ be the $d$-dimensional unit cube,
let $C(D)$ denote the space of continuous functions on $D$, 
endowed with the supremum norm.
For 
$1\le p \le\infty$, let $L_p(D)$
be the space of real-valued $p$-integrable functions, equipped with the usual norm
$$\|f\|_{L_p(D)}=\left(\int_D|f(t)|^pdt\right)^{1/p}$$ 
if $p<\infty$, and 
$$
\|f\|_{L_\infty(D)}={\rm ess\, sup}_{t\in D}|f(t)|.
$$
The Sobolev space
$W_p^r(D)$ consists of all functions $f\in L_p(D)$ such that for all 
$\a=(\a_1,\dots,\a_d)\in N_0^d$
with $|\a|:=\sum_{j=1}^d\a_j\le r$, 
the generalized partial derivative $\partial^\a f$ belongs to $L_p(D)$. The norm 
on $W_p^r(D)$ is defined as 
$$\|f\|_{W_p^r(D)}=\left(\sum_{|\a|\le r}\|\partial^\a f\|^p_{L_p(D)}\right)^{1/p}$$
if $p<\infty$, and
$$
\|f\|_{W_\infty^r(D)}=\max_{|\a|\le r}\|\partial^\a f\|_{L_\infty (D)}.
$$

We always assume that $r/d>1/p$. By the Sobolev embedding theorem (see
\cite{Ada75}, \cite{Tri95}), functions from $W_p^r(D)$ are continuous,
 and therefore function values are well-defined. Let $\mathcal{B}(W_p^r(D))$ 
 be the unit ball of
the space $W_p^r(D)$ and $J_{pq}:W_p^r(D)\to L_q(D)$ 
the embedding operator
$J_{pq}f=f\quad(f\in W_p^r(D))$. 

Now we present the main result of this paper. To emphasize the essential parts 
of the estimates we introduce the following notation. For functions 
$a,b:\N\to [0,\infty)$, we write 
$a(n)\asymp_{\log} b(n)$ if there are constants 
$c_1,c_2>0$, $n_0\in \N$, $\alpha_1,\alpha_2\in \R$ such that 
$$
c_1(\log(n+1))^{\alpha_1}b(n)\le a(n)\le c_2(\log(n+1))^{\alpha_2}b(n)
$$
for all 
$n\in \N$ with $n \ge n_0$. 
Throughout the paper $\log$ means $\log_2$. Furthermore, we often use the 
same symbol $c,c_1,\dots$ for possibly different
positive constants (also when they appear in a sequence of relations).
These constants are either absolute or may depend only on $p,q,r$ and $d$ -- 
in all statements of lemmas, propositions, etc. this is precisely described anyway 
by the order of the quantifiers.

\begin{theorem}
\label{theo:2}
Let $r,d\in\N$, $1\le p,q \le\infty$ and assume $r/d>1/p$.  
Then for $r/d\ge 2/p-2/q$
$$
 e_n^\q(J_{pq},\mathcal{B}(W_p^r(D)))\asymp_{\log} n^{-r/d},
$$
while for $r/d< 2/p-2/q$
$$
 e_n^\q(J_{pq},\mathcal{B}(W_p^r(D)))\asymp_{\log} n^{-2r/d+2/p-2/q}.
$$
\end{theorem}
Here $e_n^\q(J_{pq},\mathcal{B}(W_p^r(D)))$ is the $n$-th minimal query error,
that is, the minimal possible error among
all quantum algorithms that use at most $n$ query calls to approximate
$J_{pq}$ on $\mathcal{B}(W_p^r(D))$, in the norm of $L_q(D)$ (see \cite{Hei03a}
for the definition of $e_n^\q$). Theorem \ref{theo:2} is a direct 
consequence of 
Propositions \ref{pro:8} and \ref{pro:9}, which are stated and proved below 
and which also contain the logarithmic factors.
First we derive the upper bounds.

\begin{proposition}
\label{pro:8}
Let $r,d\in\N$, $1\le p,q \le\infty$ and assume $r/d>1/p$.  
Then there exists a constant $c>0$ such that for all $n\in\N$ with $n>2$
the following hold:
For $p < q$ and $r/d>2/p-2/q$, 
\begin{equation}
\label{XA2}
e_n^\q(J_{pq},\mathcal{B}(W_p^r(D)))
\le c n^{-r/d}(\log n)^{2/p-2/q},
\end{equation}
for $p < q$ and $r/d=2/p-2/q$,
\begin{equation}
\label{XA3}
e_n^\q(J_{pq},\mathcal{B}(W_p^r(D)))
\le c n^{-r/d}(\log n)^{4/p-4/q+1}(\log\log n)^{2/p-2/q},
\end{equation}
for
$p < q$ and $r/d<2/p-2/q$,
\begin{equation}
\label{XA4}
e_n^\q(J_{pq},\mathcal{B}(W_p^r(D)))
\le c n^{-2r/d+2/p-2/q},
\end{equation}
and for $p\ge q$,
\begin{equation}
\label{XA1}
e_n^\q(J_{pq},\mathcal{B}(W_p^r(D)))\le c n^{-r/d}.
\end{equation}
\end{proposition}
\begin{proof} We need some preparations. We  
show that the  discretization technique developed in \cite{Hei01b}, properly adapted,
works also for the approximation problem. For the sake of completeness, 
we recall also needed details  from \cite{Hei01b}.
For $l\in\N_{0}$ let 
\begin{eqnarray*}
D= \bigcup_{i=0}^{2^{dl}-1} D_{li}
\end{eqnarray*}
be the partition of $D$ into $2^{dl}$  congruent cubes of disjoint interior.
Let $s_{li}$ denote the point in $D_{li}$ with the smallest Euclidean norm.
Introduce the following  operators  $E_{li}$ and 
$R_{li}$ from $\mathcal{F}(D,\R)$, the set of all real-valued functions on $D$,
 to $\mathcal{F}(D,\R)$,
by setting for $f\in\mathcal{F}(D,\R)$ and $s\in D$
$$
(E_{li}f)(s)=f(s_{li}+2^{-l}s)
$$
and
\[
(R_{li}f)(s)=
\left\{\begin{array}{lll}
  f(2^{l}(s-s_{li})) & \mbox{if} \quad  s\in D_{li}  \\
   0 & \mbox{otherwise.}    \\
    \end{array}
\right. 
\]
Let $P$ be any operator from $C(D)$ to $L_\infty(D)$
of the form  
$$
Pf=\sum_{j=0}^{\kappa-1}  f(t_j)\phi_j\quad(f\in C(D))
$$
with $t_j\in D$ and $\phi_j\in L_{\infty}(D)$.  Assume furthermore that   
$P$ is the identity on $\mathcal{P}_{r-1}(D)$,
that is, 
\begin{equation}
\label{A1}
Pf= f \quad \mbox{for all}\quad f\in\mathcal{P}_{r-1}(D),
\end{equation}
 where
$\mathcal{P}_{r-1}(D)$ denotes the space of
polynomials on $D$ of degree not exceeding $r-1$. (For example, for $d=1$ 
one can take Lagrange interpolation of appropriate degree and
for $d>1$ its tensor product.) Since $r>d/p$, we have, by the Sobolev embedding
theorem \cite{Ada75}, \cite{Tri95}, 
$W_p^r (D)\subset C(D)$ and there is a constant $c>0$ such that for
each $f\in W_p^r (D)$
\begin{equation}
\label{E1}
\|f\|_{C(D)}\le c\|f\|_{W_p^r (D)}.
\end{equation}
It follows that
\begin{equation}
\label{A2}
\|Pf\|_{L_q(D)}\le\sum_{j=0}^{\kappa-1} |f(t_j)|\|\phi_j\|_{L_q(D)}
\le \sum_{j=0}^{\kappa-1}\|\phi_j\|_{L_q(D)} \|f\|_{C(D)}\le c\|f\|_{W_p^r (D)}
\end{equation}
(in what follows the operator $P$ will be fixed, hence 
$\sum_{j=0}^{\kappa-1}\|\phi_j\|_{L_q(D)}$ can be considered as a constant).
For $f\in W_p^r (D)$ we denote 
$$
|f|_{r,p,D}=\left(\sum_{|\alpha|=r}\|\partial^\a f\|^p_{L_p(D)}\right)^{1/p}
$$
if $p<\infty$, and
$$
|f|_{r,\infty,D}=\max_{|\a|= r}\|\partial^\a f\|_{L_\infty (D)}.
$$

Now we use Theorem 3.1.1 in \cite{Cia78}: there is a constant $c>0$ such that 
for all $f\in W_p^r (D)$
\begin{equation}
\label{B1}
\inf_{g\in\mathcal{P}_{r-1}(D)} \|f-g\|_{W_p^r (D)}
\le c|f|_{r,p,D}.
\end{equation}
By (\ref{A1}), (\ref{A2}) and (\ref{B1}),
\begin{eqnarray}
\|f-Pf\|_{L_q(D)}&\le&\inf_{g\in\mathcal{P}_{r-1}(D)}\|(f-g)-P(f-g)\|_{L_q(D)}\nonumber\\ 
&\le& c\inf_{g\in\mathcal{P}_{r-1}(D)} \|f-g\|_{W_p^r (D)}
\le c |f|_{r,p,D}\label{A4}.
\end{eqnarray}
For $l\in\N_0$ set 
$$
P_l f = \sum_{i=0}^{2^{dl}-1}R_{li} PE_{li}f
=\sum_{i=0}^{2^{dl}-1}\sum_{j=0}^{\kappa-1} f(s_{li}
+2^{-l}t_j)R_{li}\phi_j.
$$
Note that 
\begin{equation}
\label{AA7}
\|R_{li}f\|_{L_p(D)}=2^{-dl/p}\|f\|_{L_p(D)}\quad (f\in L_p(D)). 
\end{equation}
Then we have 
for $u\in\{p,q\}$ and all $f\in W_p^r (D)$, 
using (\ref{A4}) and (\ref{AA7}), 
\begin{eqnarray}
\|f-P_lf\|_{L_u(D)}
&=&\|\sum_{i=0}^{2^{dl}-1}(R_{li}E_{li}f-R_{li}PE_{li}f)\|_{L_u(D)}
\nonumber\\
&=&\left(\sum_{i=0}^{2^{dl}-1}\|(R_{li}(E_{li}f-PE_{li}f)\|^u_{L_u(D)}\right)^{1/u}
\nonumber\\
&=&\left(2^{-dl}\sum_{i=0}^{2^{dl}-1}\|E_{li}f-PE_{li}f\|^u_{L_u(D)}\right)^{1/u}
\nonumber\\
&\le&c\left(2^{-dl}\sum_{i=0}^{2^{dl}-1}|E_{li}f|^u_{r,p,D}\right)^{1/u}
\nonumber\\
&\le& c\,2^{\max(1/p-1/u,0)dl}\left(2^{-dl}\sum_{i=0}^{2^{dl}-1}|E_{li}f|^p_{r,p,D}\right)^{1/p}\nonumber
\end{eqnarray}
and
\begin{eqnarray}
\left(2^{-dl}\sum_{i=0}^{2^{dl}-1}|E_{li}f|_{r,p,D}^p\right)^{1/p}
&=&\left(2^{-dl}\sum_{i=0}^{2^{dl}-1}\sum_{|\alpha|=r}
\int_D |\partial^\alpha f(s_{li}+2^{-l}t)|^p\,dt\right)^{1/p}\nonumber\\
&=&2^{-rl}\left(\sum_{i=0}^{2^{dl}-1}\sum_{|\alpha|=r}
\int_{D_{li}} |\partial^\alpha f(t)|^p\,dt\right)^{1/p}\nonumber\\
&=&\,2^{-rl}|f|_{r,p,D}\le \,2^{-rl}\|f\|_{W_p^r (D)}\label{A5}
\end{eqnarray}
(with the usual modifications for $u=\infty$ or $p=\infty$).
Consequently,
\begin{eqnarray}
\|f-P_lf\|_{L_u(D)}&\le& c\,2^{-rl+\max(1/p-1/u,0)dl}|f|_{r,p,D}\nonumber\\
&\le& c\,2^{-rl+\max(1/p-1/u,0)dl}\|f\|_{W_p^r (D)}.\label{B2}
\end{eqnarray}

Similarly to \cite{Hei01b}, we first approximate $f$ by 
$P_{l^*}f$ for some $l^*$, giving the desired
precision, but using a number
of function values much larger than $n$. This $P_{l^*}$, in turn, will be split 
into the sum of
a single operator $P_{l_0}$, with number of function values of the order $n$, 
which we compute
classically, and a hierarchy of operators 
$P_l'$ $(l=l_0,\dots, l^*-1)$. We will show that the approximation of the $P'_l$ 
reduces to the 
approximation of appropriately scaled embedding operators 
$J_{pq}^{N_l}: L_p^{N_l}\to L_q^{N_l}$ for suitable
$N_l$. This enables us to apply the results of \cite{Hei03a}.
Define
\begin{eqnarray}
\lefteqn{P'f:=(P_1-P_0)f}\nonumber\\
&=&\sum_{i=0}^{2^{d}-1}\sum_{j=0}^{\kappa-1}  
f(s_{1,i}+2^{-1}t_j)R_{1,i}\phi_j
-\sum_{j=0}^{\kappa-1}  f(t_j)\phi_j\label{A6},
\end{eqnarray}
which can be written as
$$
P'f=\sum_{j=0}^{\kappa'-1}\left(\sum_{k=0}^{\kappa''-1}a_{jk}f(t'_{jk})\right) \psi_j,
$$
with 
\begin{equation}
\label{E2}
\kappa',\kappa''\le\kappa (2^d+1),
\end{equation}
$a_{jk}\in\R,\quad t'_{jk}\in D \quad (j=0,\dots,\kappa'-1,\, k=0,\dots,\kappa''-1)$,
 and a linearly independent system  
$(\psi_j)_{j=0}^{\kappa'-1}\subset L_\infty(D)$.
The linear independence implies that for $u\in \{p,q\}$ there are constants
$c_1,c_2>0$ such that
for all $\a_j\in\R$ $(j=0,\dots,\kappa'-1)$
\begin{equation}
\label{AA1}
c_1\|(\a_j)_{j=0}^{\kappa'-1}\|_{L_u^{\kappa'}}
\le \|\sum_{j=0}^{\kappa'-1}\a_j\psi_j\|_{L_u(D)}
\le c_2\|(\a_j)_{j=0}^{\kappa'-1}\|_{L_u^{\kappa'}}.
\end{equation}
Put 
\begin{equation}
\label{AA2}
\psi_{lij}=R_{li}\psi_j,
\end{equation}
and let
$$
\Pi_{l}={\rm span}\left\{\psi_{lij}\;:\;i=0,\dots,2^{dl}-1,\;
j=0,\dots,\kappa'-1\right\}\subseteq L_\infty(D).
$$ 
By the disjointness of the interiors of the $D_{li}$ and by (\ref{AA7})
we have for $\a_{ij}\in\R$ $(i=0,\dots, 2^{dl}-1, j=0,\dots,\kappa'-1)$
and $1\le u< \infty$
\begin{eqnarray}
\lefteqn{
\|\sum_{i=0}^{2^{dl}-1}\sum_{j=0}^{\kappa'-1}\a_{ij}\psi_{lij}\|^u_{L_u(D)}}\nonumber\\
&=&\sum_{i=0}^{2^{dl}-1}\|\sum_{j=0}^{\kappa'-1}\a_{ij}\psi_{lij}\|^u_{L_u(D)}
=\sum_{i=0}^{2^{dl}-1}\|R_{li}\sum_{j=0}^{\kappa'-1}\a_{ij}\psi_j\|^u_{L_u(D)}\nonumber\\
&=&2^{-dl}\sum_{i=0}^{2^{dl}-1}\|\sum_{j=0}^{\kappa'-1}\a_{ij}\psi_j\|^u_{L_u(D)}.
\label{AA4}
\end{eqnarray}
Let $N_{l}=\kappa'\,2^{dl}$. Then
\begin{equation}
\label{AA7a}
2^{-dl}\sum_{i=0}^{2^{dl}-1}\|(\a_{ij})_{j=0}^{\kappa'-1}\|^u_{L_u^{\kappa'}}
=\|(\a_{ij})\|^u_{L_u^{N_l}},
\end{equation}
where $(\a_{ij})$ stands for $(\a_{ij})_{i=0,j=0}^{2^{dl}-1,\kappa'-1}$.
Combining (\ref{AA4}), (\ref{AA7a}) and (\ref{AA1}), we get
\begin{equation}
\label{AA3}
c_1\|(\a_{ij})\|_{L_u^{N_l}}
\le\|\sum_{i=0}^{2^{dl}-1}\sum_{j=0}^{\kappa'-1}\a_{ij}\psi_{lij}\|_{L_u(D)}
\le c_2\|(\a_{ij})\|_{L_u^{N_l}}.
\end{equation}
Relation (\ref{AA3}) holds also for $u=\infty$, which 
can be proved with the usual modifications in the
reasoning above.
Define the operator $T_l:\Pi_l\to \R^{N_l}$ by 
\begin{equation}
\label{AB2}
T_l\sum_{i=0}^{2^{dl}-1}\sum_{j=0}^{\kappa'-1}\a_{ij}\psi_{lij} = (\a_{ij}).
\end{equation}
It follows from (\ref{AA3}) that for $f\in \Pi_l$,
\begin{equation}
\label{AB3}
\|T_l f\|_{L_u^{N_l}}\le c_1^{-1}\|f\|_{L_u(D)},
\end{equation}
and for $g\in L_u^{N_l}$,
\begin{equation}
\label{AB3a}
\|T_l^{-1}g\|_{L_u(D)}\le c_2\|g\|_{L_u^{N_l}}.
\end{equation}
For $l\in\N_0$ and $f\in C(D)$ set 
\begin{eqnarray}
P'_{li}f&=&R_{li}P'E_{li}f=
\sum_{j=0}^{\kappa'-1} \sum_{k=0}^{\kappa''-1}
a_{jk} f(s_{li}+2^{-l}t_{jk}')\psi_{lij}
\label{B5},\\
P'_{l} &=& \sum_{i=0}^{2^{dl}-1} P'_{li}.\label{B6}
\end{eqnarray}
It is readily verified that
$$
P_{l+1}=\sum_{i=0}^{2^{dl}-1}R_{li}P_1 E_{li}.
$$
and therefore 
\begin{eqnarray}
P_{l+1}-P_l&=&\sum_{i=0}^{2^{dl}-1}R_{li}(P_1 E_{li} -P_0 E_{li})\nonumber\\
&=&\sum_{i=0}^{2^{dl}-1}P'_{li}=P'_l.\label{B4}
\end{eqnarray}
From (\ref{AA7}), (\ref{B2}) with $u=p$ and (\ref{A5}), we get 
\begin{eqnarray}
\lefteqn{\|P'_lf\|_{L_p(D)}}\nonumber\\
&=&\left(\sum_{i=0}^{2^{dl}-1}\|R_{li}P'E_{li}f\|_{L_p(D)}^p\right)^{1/p}
\nonumber\\
&=&
\left(2^{-dl}\sum_{i=0}^{2^{dl}-1}\|P_1 E_{li}f -P_0E_{li}f \|_{L_p(D)}^p
\right)^{1/p}\nonumber\\
&\le&\left(2^{-dl}\sum_{i=0}^{2^{dl}-1}\left(\|E_{li}f-P_1 E_{li}f\|_{L_p(D)}
+\|E_{li} f-P_0 E_{li} f\|_{L_p(D)}\right)^p\right)^{1/p}\nonumber\\
&\le&c\left(2^{-dl}\sum_{i=0}^{2^{dl}-1}|E_{li}f|_{r,p,D}^p\right)^{1/p}
\le c\,2^{-rl}\|f\|_{W_p^r (D)}.\label{B3}
\end{eqnarray}
We define operators $U_l:W_p^r(D)\to L_p^{N_l}$ by 
\begin{equation}
\label{W5}
U_l=T_l P'_l
\end{equation}
and $V_l:L_q^{N_l}\to L_q(D)$ by
\begin{equation}
\label{W6}
V_l=T_l^{-1}.
\end{equation}
Then clearly
\begin{equation}
\label{W7}
V_lJ_{pq}^{N_l}U_l=P'_l,
\end{equation}
moreover, by (\ref{B3}) and (\ref{AB3}) for $u=p$,
\begin{equation}
\label{W8}
\|U_lf\|_{L_p^{N_l}}\le c2^{-rl}\|f\|_{W_p^r(D)}\quad (f\in W_p^r(D))
\end{equation}
and, by (\ref{AB3a}) for $u=q$
\begin{equation}
\label{W9}
\|V_lg\|_{L_q(D)}\le c\|g\|_{L_q^{N_l}}\quad (g\in L_q^{N_l}).
\end{equation}

Now we are ready to derive the upper bounds.
It obviously suffices to prove them for 
\begin{equation}
\label{L1}
n\ge \max(\kappa,5).
\end{equation}
Define 
\begin{equation}
\label{D1}
l_0=\lfloor d^{-1}\log(n/\kappa)\rfloor.
\end{equation}
Then $l_0\ge 0$. Furthermore, let 
\begin{equation}
\label{H9}
l^*=\left\{\begin{array}{lll}
  l_0 & \mbox{if} \quad p\ge q   \\
  2l_0 & \mbox{if} \quad p<q.   \\
    \end{array}
\right.
\end{equation}
 By (\ref{B4}),
\begin{equation}\label{C1}
P_{l^*}=P_{l_0}+\sum_{l=l_0}^{l^*-1}P'_l.
\end{equation}
In the sequel we consider the $P_l$ and $P'_l$ as operators from 
$W_p^r(D)$ to $L_q(D)$. 
By (\ref{D1}), $\kappa\,2^{dl_0}\le n$, hence
\begin{equation}
\label{H3}
e_{n}^\q(P_{l_0},\mathcal{B}(W_p^r(D)),0)=0
\end{equation}
(this just means that with $\kappa\,2^{dl_0}$ queries
we can compute $P_{l_0}$, classically, or, more precisely, up to any precision by  
simulating the classical computation on a suitable number of qubits).
Let $\nu_l,n_l\in \N$ $(l=l_0,\dots,l^*-1)$ be natural numbers which will 
be specified later on, and which will be assumed to  
satisfy
\begin{equation}
\label{I2}
\sum_{l=l_0}^{l^*-1}e^{-\nu_l/8}\le \frac{1}{4}.
\end{equation}
Put
\begin{equation}
\label{C7}
\wt{n}=n+2\kappa''\sum_{l=l_0}^{l^*-1}\nu_l n_l
\end{equation}
(if $l^*=l_0$, we do not define the numbers $\nu_l,n_l$ and put $\wt{n}=n$).
From (\ref{B2}) above with $u=q$ and Lemma 6(i) of \cite{Hei01}, we get
\begin{eqnarray}
\label{H2}
\lefteqn{e_{\wt{n}}^\q(J_{pq},\mathcal{B}(W_p^r(D)))}\nonumber\\
&\le&
\sup_{f\in \mathcal{B}(W_p^r(D))}\|J_{pq}f-P_{l^*}f\|_{L_q(D)}
+e_{\wt{n}}^\q(P_{l^*},\mathcal{B}(W_p^r(D)))\nonumber\\
&\le &c\,2^{-rl^*+\max(1/p-1/q,0)dl^*}+e_{\wt{n}}^\q(P_{l^*},\mathcal{B}(W_p^r(D))).
\end{eqnarray}
The upper bound for the case $p\ge q$ follows directly from (\ref{H2}) and 
(\ref{H3}), since in this case  $l^*=l_0$ and $\wt{n}=n$
(this is the trivial case where the optimal rate is already attained
by a classical algorithm). 

In the rest of the proof we assume $p<q$.
By Lemma 3 of \cite{Hei03a} and (\ref{H3}),
\begin{eqnarray}
\label{H4}
\lefteqn{e_{\wt{n}}^\q(P_{l^*},\mathcal{B}(W_p^r(D)))}\nonumber\\
&\le&
e_n^\q(P_{l_0},\mathcal{B}(W_p^r(D)),0)+e_{\wt{n}-n}^\q(P_{l^*}-P_{l_0},\mathcal{B}(W_p^r(D)))
\nonumber\\
&=& e_{\wt{n}-n}^\q(P_{l^*}-P_{l_0},\mathcal{B}(W_p^r(D))).
\end{eqnarray}
From (\ref{C1}), (\ref{C7}), Corollary 3 of \cite{Hei03a}, and  (\ref{I2}) we get
\begin{eqnarray}
\label{H5}
e_{\wt{n}-n}^\q(P_{l^*}-P_{l_0},\mathcal{B}(W_p^r(D)))&=&
e_{2\kappa''\sum_{l=l_0}^{l^*-1}\nu_l n_l}^\q\left(\sum_{l=l_0}^{l^*-1}P'_l,
\mathcal{B}(W_p^r(D))\right)\nonumber\\
&\le&2\sum_{l=l_0}^{l^*-1} e_{2\kappa''n_l}^\q(P'_l,\mathcal{B}(W_p^r(D))).
\end{eqnarray}
Using Lemma 2 of \cite{Hei03a},  (\ref{W7}), and (\ref{W9}) we obtain
\begin{eqnarray}
e_{2\kappa''n_l}^\q(P'_l,\mathcal{B}(W_p^r(D)))
&=&
e_{2\kappa''n_l}^\q(V_l J_{pq}^{N_l} U_l,\mathcal{B}(W_p^r(D)))\nonumber\\
&\le&c\,e_{2\kappa''n_l}^\q(J_{pq}^{N_l} U_l,\mathcal{B}(W_p^r(D))).
\label{H6}
\end{eqnarray}
Joining relations (\ref{H2})--(\ref{H6}),
we infer
\begin{eqnarray}
\lefteqn{e_{\wt{n}}^\q(J_{pq},\mathcal{B}(W_p^r(D)))}\nonumber\\
&\le& c\,2^{-rl^*+(1/p-1/q)dl^*}+c\sum_{l=l_0}^{l^*-1} e_{2\kappa''n_l}^\q(J_{pq}^{N_l}U_l,
\mathcal{B}(W_p^r(D))).
\label{H8}\end{eqnarray}
In a further reduction one would like to remove the $U_l$ in the last relation.
This could be done on the basis of Corollary 1 of \cite{Hei01b}, if the $U_l$ were 
of the required form, which is not the case. Instead, we shall
approximate the $U_l$ by appropriate mappings $\Gamma_l:\mathcal{B}(W_p^r(D))\to L_p^{N_l}
\quad (l_0\le l<l^*)$.
Note that by (\ref{AB2}), (\ref{B5}), (\ref{B6}), and (\ref{W5}),
\begin{equation}
\label{W10}
U_lf(i,j)=\sum_{k=0}^{\kappa''-1}
a_{jk} f(s_{li}+2^{-l}t_{jk}').
\end{equation}
Fix an $m^*\in \N$ with 
\begin{equation}
\label{G1}
2^{-m^*/2}\le (l^*+1)^{-1}2^{-rl^*}
\end{equation}
and
\begin{equation}
\label{G2}
2^{m^*/2-1}\ge c,
\end{equation}
where $c$ is the constant from (\ref{E1}).
Hence,
\begin{equation}
\label{E3}
\|f\|_{C(D)}\le 2^{m^*/2-1}\quad \mbox{for}\quad f\in\mathcal{B}(W_p^r(D)).
\end{equation}
Define
$\b:\R\to\Z[0,2^{m^*})$ for $z\in\R$ by
\begin{equation}\label{N1}
\b(z)=
\left\{\begin{array}{lll}
   0& \mbox{if} \quad z <-2^{m^*/2-1} \\
   \lfloor 2^{m^*/2}(z+2^{m^*/2-1})\rfloor       & \mbox{if} \quad  
   -2^{m^*/2-1}\le z <2^{m^*/2-1}\\
   2^{m^*}-1& \mbox{if} \quad z\ge 2^{m^*/2-1} 
   \end{array}
   \right.
\end{equation}
and $\gamma:\Z[0,2^{m^*})\to\R$  for $y\in\Z[0,2^{m^*})$ as 
\begin{equation}
\label{N2}
\g(y)=2^{-m^*/2}y-2^{m^*/2-1}.
\end{equation}
Then we have for $-2^{m^*/2-1}\le z\le 2^{m^*/2-1}$,
\begin{equation}
\label{E4}
\g(\b(z))\le z\le \g(\b(z))+2^{-m^*/2}.
\end{equation}
Define $\eta_{lk}:\Z[0,N_l)\to D\quad (k=0,\dots,\kappa''-1)$ by
$$
\eta_{lk}(i,j)=s_{li}+2^{-l}t_{jk}'\quad (0\le i \le 2^{dl}-1,\;0\le j \le \kappa'-1)
$$
(here $\Z[0,N_l)$ stands for $\{0,1,\dots,N_l-1\}$ and we 
identify $\Z[0,N_l)$ with $\Z[0,2^{dl})\times\Z[0,\kappa')$).
Next let $\rho_l:\Z[0,N_l)\times\Z[0,2^{m^*})^{\kappa''}\to\R$ be given by
$$
\rho_l((i,j),y_0,\dots,y_{\kappa''-1})=\sum_{k=0}^{\kappa''-1}a_{jk}\g(y_k).
$$
 Finally, we define $\Gamma_l:\mathcal{B}(W_p^r(D))\to L_p^{N_l}$ by setting
$$
\Gamma_l(f)(i,j)=\rho_l((i,j),(\b\circ f\circ\eta_{lk}(i,j))_{k=0}^{\kappa''-1}).
$$
for $f\in \mathcal{B}(W_p^r(D))$.
Note that $\Gamma_l$ is of the form (4) of \cite{Hei01b} needed to apply Corollary 
1 of that paper, which we will do later on.
We have 
$$
\Gamma_l(f)(i,j)=\sum_{k=0}^{\kappa''-1}a_{jk}\g(\b(f(s_{li}+2^{-l}t_{jk}'))),
$$
hence, by (\ref{W10}), (\ref{E3}), (\ref{E4}), and (\ref{G1})
\begin{eqnarray}
\label{H1}
\lefteqn{|(U_l f)(i,j)-\Gamma_l(f)(i,j)|}\nonumber\\
&\le& \sum_{k=0}^{\kappa''-1}|a_{jk}|
|f(s_{li}+2^{-l}t_{jk}')-\g(\b(f(s_{li}+2^{-l}t_{jk}')))|\nonumber\\
&\le& 2^{-m^*/2}\sum_{k=0}^{\kappa''-1}|a_{jk}|\le c\,2^{-m^*/2}\le c(l^*+1)^{-1}2^{-rl^*},
\end{eqnarray}
and therefore, for all $f\in \mathcal{B}(W_p^r(D))$ and $u\in \{p,q\}$,
\begin{eqnarray}
\lefteqn{
\|U_l f-\Gamma_l(f)\|_{L_u^{N_l}}}\nonumber\\
&=&\left((\kappa')^{-1} 2^{-dl}\sum_{i=0}^{2^{dl}-1}\sum_{j=0}^{\kappa'-1}
|(U_l f)(i,j)-\Gamma_l(f)(i,j)|
^u\right)^{1/u}\nonumber\\
&\le& c(l^*+1)^{-1}2^{-rl^*}. \label{E5}
\end{eqnarray}
Moreover, by (\ref{W8}) and (\ref{E5}) with $u=p$, 
\begin{eqnarray*}
\|\Gamma_l(f)\|_{L_p^{N_l}}&\le& \|U_lf\|_{L_p^{N_l}}
+\|\Gamma_l(f)-U_lf\|_{L_p^{N_l}}
\le c\,2^{-rl}.
\end{eqnarray*}
Consequently, 
\begin{equation}
\label{C4}
\Gamma_l (\mathcal{B}(W_p^r(D)))\subseteq c\,2^{-rl}\mathcal{B}(L_p^{N_l}).
\end{equation}
From (\ref{E5}) with $u=q$ and Lemma 6(i) of \cite{Hei01}
\begin{eqnarray}
\lefteqn{
e_{2\kappa''n_l}^\q(J_{pq}^{N_l} U_l,\mathcal{B}(W_p^r(D)))
}\nonumber\\
&\le&c(l^*+1)^{-1}2^{-rl^*}+e_{2\kappa''n_l}^\q(J_{pq}^{N_l}\Gamma_l,\mathcal{B}(W_p^r(D))).
\label{H6a}
\end{eqnarray}
Corollary 1 of \cite{Hei01b}, relation (\ref{C4}) above and 
Lemma 6(iii) of \cite{Hei01} give
\begin{eqnarray}
\label{H7}
e_{2\kappa''n_l}^\q(J_{pq}^{N_l}\Gamma_l,\mathcal{B}(W_p^r(D)))
&\le& e_{n_l}^\q(J_{pq}^{N_l},c\,2^{-rl}\mathcal{B}(L_p^{N_l}))\nonumber\\
&=& c\,2^{-rl}e_{n_l}^\q(J_{pq}^{N_l},\mathcal{B}(L_p^{N_l})).
\end{eqnarray}
From (\ref{H6a}), (\ref{H7}) and (\ref{H8}),
we conclude
\begin{eqnarray}\label{H8a1}
\lefteqn{e_{\wt{n}}^\q(J_{pq},\mathcal{B}(W_p^r(D)))}\nonumber\\
&\le& c\,2^{-rl^*+(\frac{1}{p}-\frac{1}{q})dl^*}+
c\sum_{l=l_0}^{l^*-1} 2^{-rl}e_{n_l}^\q(J_{pq}^{N_l},\mathcal{B}(L_p^{N_l})).
\end{eqnarray}
Thus we reached the desired reduction and can now exploit the results 
for the finite dimensional case: By  
Proposition 2 of \cite{Hei03a} and (\ref{H8a1}),
\begin{eqnarray}
\lefteqn{e_{\wt{n}}^\q(J_{pq},\mathcal{B}(W_p^r(D)))}\nonumber\\
&\le&
c\,2^{-rl^*+(\frac{1}{p}-\frac{1}{q})dl^*}\nonumber\\
&& +c\sum_{l=l_0}^{l^*-1} 2^{-rl}
n_l^{-(\frac{2}{p}-\frac{2}{q})}N_l^{\frac{2}{p}-\frac{2}{q}}
(\log(n_l/\sqrt{N_l}+2))^{\frac{2}{p}-\frac{2}{q}}.\label{H8a} 
\end{eqnarray}
Recall that we consider the case $p <q$. First we assume 
\begin{equation}
\label{AG1}
\frac{r}{d}>\frac{2}{p}-\frac{2}{q}.
\end{equation}
Fix any $\delta>0$ such that
\begin{equation}
\label{N5}
r>\left(\frac{2}{p}-\frac{2}{q}\right)(d+\delta)
\end{equation}
and put for $l=l_0,\dots,l^*-1$
\begin{equation}
\label{I1}
 n_l=\left\lceil 2^{dl_0-\delta(l-l_0)}\right\rceil,
\end{equation}
\begin{equation}
\label{I1A} 
 \nu_l=\left\lceil 8(2\ln(l-l_0+1)+ \ln 8)\right\rceil. 
\end{equation}
It follows from (\ref{I1A}) that 
\begin{equation}
\label{I2D}
\sum_{l=l_0}^{l^*-1}e^{-\nu_l/8}\le \frac{1}{8}\sum_{l=l_0}^{l^*-1}(l-l_0+1)^{-2}<\frac{1}{4},
\end{equation}
so assumption (\ref{I2}) is satisfied.
By (\ref{C7}), (\ref{I1A}), (\ref{D1}), and (\ref{H9}),
\begin{eqnarray}
\label{I3}
\wt{n}&\le& n+2\kappa''\sum_{l=l_0}^{l^*-1} 
\left\lceil 8(2\ln(l-l_0+1)+ \ln 8)\right\rceil 
\left\lceil2^{dl_0-\delta(l-l_0)}\right\rceil\nonumber\\
&\le& c\,2^{dl_0} \le cn.
\end{eqnarray}
It follows from (\ref{H8a}), 
(\ref{L1})--(\ref{H9}), (\ref{N5}), and (\ref{I1}) that
\begin{eqnarray}
&&e_{\wt{n}}^\q(J_{pq},\mathcal{B}(W_p^r(D)))\nonumber\\
&\le& c\,2^{-rl^*/2}+c\sum_{l=l_0}^{l^*-1} 
2^{-rl-(\frac{2}{p}-\frac{2}{q})dl_0+(\frac{2}{p}-\frac{2}{q})\delta(l-l_0)
+(\frac{2}{p}-\frac{2}{q})dl}
(l_0+1)^{\frac{2}{p}-\frac{2}{q}}\nonumber\\
&=& c\,2^{-rl_0}+c\,2^{-rl_0}(l_0+1)^{\frac{2}{p}-\frac{2}{q}}
\sum_{l=l_0}^{l^*-1}2^{(-r+(\frac{2}{p}-\frac{2}{q})(d+\delta))(l-l_0)}\nonumber\\
&\le& c\,2^{-rl_0}(l_0+1)^{\frac{2}{p}-\frac{2}{q}}
\le cn^{-\frac{r}{d}}(\log n)^{\frac{2}{p}-\frac{2}{q}}.\label{V1}
\end{eqnarray}
Then (\ref{XA2}) follows by an obvious scaling 
from (\ref{I3}) and (\ref{V1}).
Next we assume 
$$
\frac{r}{d}<\frac{2}{p}-\frac{2}{q}.
$$
Here we take any $\delta>$ with 
\begin{equation}
\label{N5a}
r<\left(\frac{2}{p}-\frac{2}{q}\right)(d-\delta)
\end{equation}
and put for $l=l_0,\dots,l^*-1$
\begin{equation}
\label{I1B}
 n_l=\left\lceil 2^{dl_0-\delta(l^*-l)}\right\rceil,
\end{equation}
\begin{equation}
\label{I1AB} 
 \nu_l=\left\lceil 8(2\ln(l^*-l)+ \ln 8)\right\rceil. 
\end{equation}
From (\ref{I1AB}) we see that (\ref{I2}) is satisfied, again:
\begin{equation}
\label{I2C}
\sum_{l=l_0}^{l^*-1}e^{-\nu_l/8}\le \frac{1}{8}\sum_{l=l_0}^{l^*-1}(l^*-l)^{-2}<\frac{1}{4}.
\end{equation}
By (\ref{C7}), (\ref{I1B}), (\ref{I1AB}), (\ref{D1}), and (\ref{H9}),
\begin{eqnarray}
\label{I3B}
\wt{n}&\le& n+2\kappa''\sum_{l=l_0}^{l^*-1} 
\left\lceil 8(2\ln(l^*-l)+ \ln 8)\right\rceil 
\left\lceil2^{dl_0-\delta(l^*-l)}\right\rceil\nonumber\\
&\le& c\,2^{dl_0} \le cn.
\end{eqnarray}
Using (\ref{H8a}), 
(\ref{L1})--(\ref{H9}),  (\ref{N5a}), and (\ref{I1B}), we get
\begin{eqnarray*}
&&e_{\wt{n}}^\q(J_{pq},\mathcal{B}(W_p^r(D)))\\
&\le& c\,2^{(-\frac{2r}{d}+\frac{2}{p}-\frac{2}{q})dl_0}\\
&&+c\sum_{l=l_0}^{l^*-1} 
2^{-rl-(\frac{2}{p}-\frac{2}{q})dl_0+(\frac{2}{p}-\frac{2}{q})\delta(l^*-l)
+(\frac{2}{p}-\frac{2}{q})dl}
(l^*-l+1)^{\frac{2}{p}-\frac{2}{q}}\\
&=& c\,2^{(-\frac{2r}{d}+\frac{2}{p}-\frac{2}{q})dl_0}
\\
&&+c\,2^{-rl^*+(\frac{2}{p}-\frac{2}{q})dl_0}
\sum_{l=l_0}^{l^*-1}(l^*-l+1)^{\frac{2}{p}-\frac{2}{q}}
2^{(r-(\frac{2}{p}-\frac{2}{q})(d-\delta))(l^*-l)}\\
&\le& c\,2^{(-\frac{2r}{d}+\frac{2}{p}-\frac{2}{q})dl_0}
\le cn^{-\frac{2r}{d}+\frac{2}{p}-\frac{2}{q}},
\end{eqnarray*}
which, together with (\ref{I3B}), gives (\ref{XA4}).
Finally we consider the case
$$
\frac{r}{d}=\frac{2}{p}-\frac{2}{q}.
$$
Here we put for $l=l_0,\dots,l^*-1$
\begin{equation}
\label{I1C}
 n_l=\left\lceil 2^{dl_0}(l_0+1)^{-1}(\ln(l_0+2))^{-1}\right\rceil,
\end{equation}
and 
\begin{equation}
\label{I1AC} 
 \nu_l=\left\lceil 8(\ln(l_0+2)+\ln 4)\right\rceil. 
\end{equation}
This way, relation (\ref{I2}) is valid:
\begin{equation}
\label{I2E}
\sum_{l=l_0}^{l^*-1}e^{-\nu_l/8}\le \frac{1}{4}\sum_{l=l_0}^{l^*-1}(l_0+2)^{-1}<\frac{1}{4}.
\end{equation}
From (\ref{C7}), (\ref{I1C}), (\ref{I1AC}), (\ref{D1}), and (\ref{H9}), we get
\begin{eqnarray}
\label{I3C}
\wt{n}&\le& n+2\kappa''\sum_{l=l_0}^{l^*-1} 
\left\lceil 8(\ln(l_0+2)+\ln 4) \right\rceil 
\left\lceil 2^{dl_0}(l_0+1)^{-1}(\ln(l_0+2))^{-1}\right\rceil\nonumber\\
&\le& c\,2^{dl_0} \le cn.
\end{eqnarray}
By the help of (\ref{H8a}), 
(\ref{L1})--(\ref{H9}), and (\ref{I1C}) we conclude that
\begin{eqnarray*}
&&e_{\wt{n}}^\q(J_{pq},\mathcal{B}(W_p^r(D)))\\
&\le& c\,2^{-rl^*/2}+c\sum_{l=l_0}^{l^*-1} 
2^{-rl-(\frac{2}{p}-\frac{2}{q})dl_0+(\frac{2}{p}-\frac{2}{q})dl}
(l_0+1)^{\frac{4}{p}-\frac{4}{q}}(\log(l_0+2))^{\frac{2}{p}-\frac{2}{q}}\\
&=& c\,2^{-rl_0}+c\,(l_0+1)^{\frac{4}{p}-\frac{4}{q}}
(\log(l_0+2))^{\frac{2}{p}-\frac{2}{q}}\sum_{l=l_0}^{l^*-1}2^{-rl_0}\\
&\le& c\,2^{-rl_0}(l_0+1)^{\frac{4}{p}-\frac{4}{q}+1}(\log(l_0+2))^{\frac{2}{p}-\frac{2}{q}}
\\
&\le& cn^{-\frac{r}{d}}(\log n)^{\frac{4}{p}-\frac{4}{q}+1}
(\log\log n)^{\frac{2}{p}-\frac{2}{q}}.
\end{eqnarray*}
This shows (\ref{XA4}) and completes the proof.
\end{proof}

\begin{proposition}
\label{pro:9}
Let $r,d\in\N$, $1\le p,q \le\infty$, and suppose $r/d>1/p$.  
Then there exists a  constant $c>0$ such that for all $n\in\N$ with $n>4$
the following holds:
First assume  $r/d>2/p-2/q$. If $2<q\le \infty$, then 
\begin{equation}
\label{XA1a}
e_n^\q(J_{pq},\mathcal{B}(W_p^r(D)))\ge c n^{-r/d},
\end{equation}
if $q=2$, then 
\begin{equation}
\label{XA2a}
e_n^\q(J_{p,2},\mathcal{B}(W_p^r(D)))
\ge c n^{-r/d}(\log\log n)^{-3/2}(\log\log\log n)^{-1},
\end{equation}
and if $1\le q<2$, then 
\begin{equation}
\label{XA3a}
e_n^\q(J_{pq},\mathcal{B}(W_p^r(D)))
\ge c n^{-r/d}(\log n)^{-2/q+1}.
\end{equation}
Now assume $r/d\le 2/p-2/q$. Then
\begin{equation}
\label{XA4a}
e_n^\q(J_{pq},\mathcal{B}(W_p^r(D)))
\ge c n^{-2r/d+2/p-2/q}.
\end{equation}
\end{proposition}

\begin{proof}
Let $\psi$ be a $C^\infty$ function on $\R^d$ with 
$$
{\rm supp}\, \psi\subset (0,1)^d, \quad \sigma_1:=\int_{D}\psi(t)\,dt>0,
$$
and denote $\|\psi\|_{W_p^r(D)}=\sigma_2$.
Let $n\in \N$, $k\in \N_0$, and $N=2^{dk}$.
Let $R_{ki}$ and $D_{ki}$ be as defined in the beginning of the proof of 
Proposition \ref{pro:8}.
Set
$$
\psi_i(t)=R_{ki}\psi \quad (i=0,\dots,N-1).
$$
We have
\begin{equation}
\label{G7}
\int_{D_{ki}}\psi_i(t)\,dt=\sigma_1 N^{-1}
\end{equation}
and
$$
\|\psi_i\|_{W_p^r(D)}\le 2^{(r-d/p)k}\|\psi\|_{W_p^r(D)}=\sigma_2 2^{(r-d/p)k}.
$$
Consequently, taking into account the disjointness of the supports
of the $\psi_i$, for all $a_i\in\R\quad (i=0,\dots,N-1)$,
\begin{equation}
\label{N7}
\Big\|\sum_{i=0}^{N-1}a_i\psi_i\Big\|_{W_p^r(D)}
=\left(\sum_{i=0}^{N-1}|a_i|^p\|\psi_i\|_{W_p^r(D)}^p\right)^{1/p}
\le \sigma_2 2^{rk}\big\|(a_i)_{i=0}^{N-1}\big\|_{L_p^N} 
\end{equation}
(which holds also for $p=\infty$).
Fix any $m^*\in\N$ with 
\begin{equation}
\label{N8}
m^*/2-1\ge dk/p.
\end{equation}
Let $\b:\R\to\Z[0,2^{m^*})$ and $\gamma:\Z[0,2^{m^*})\to\R$ be defined as in
(\ref{N1}) and (\ref{N2}). For $f\in \mathcal{B}(L_p^N)$ we have 
$$
|f(i)|\le N^{1/p}=2^{dk/p}\le 2^{m^*/2-1}.
$$
Hence, by (\ref{E4}), for $0\le i <N$,
\begin{equation}
\label{N9}
\g(\b(f(i)))\le f(i)\le \g(\b(f(i)))+2^{-m^*/2}.
\end{equation}
Define 
$$
\Gamma:\mathcal{B}(L_p^N)\to W_p^r(D) \quad\mbox{by}
\quad\Gamma(f)=\sum_{i=0}^{N-1}\g\circ\b\circ f(i)\,\psi_i.
$$
By (\ref{N7}) and (\ref{N9}), for $f\in \mathcal{B}(L_p^N)$,
\begin{eqnarray}
\|\Gamma(f)\|_{W_p^r(D)}&\le& \sigma_2 2^{rk}\|\g\circ\b\circ f\|_{L_p^N}\nonumber\\ 
&\le& \sigma_2 2^{rk}\left(\|f\|_{L_p^N}+\|f-\g\circ\b\circ f\|_{L_p^N}\right)\nonumber\\
&\le&\sigma_2 2^{rk}\left(1+2^{-m^*/2}\right)\label{EF1}.
\end{eqnarray}
Define $\Phi:L_q(D)\to L_q^N$ by 
$$
(\Phi f)(i)=N\int_{D_{ki}}f(t)dt.
$$
It follows from (\ref{G7}) that 
\begin{equation}
\label{AH2}
\Phi\psi_i =\sigma_1 e_i,
\end{equation}
where $e_i$ denotes the $i$-th unit vector in $L_p^N$. Moreover,
by H\"older's inequality, 
\begin{eqnarray*}
\|\Phi f\|_{L_q^N}^q&=&N^{q-1}\sum_{i=0}^{N-1}\left|\int_{D_{ki}}f(t)dt\right|^q\\
&\le& N^{q-1}\sum_{i=0}^{N-1}\int_{D_{ki}}|f(t)|^qdt\,|D_{ki}|^{q-1}=\|f\|_{L_q(D)}^q.
\end{eqnarray*}
Thus, since $\Phi$ is linear,
\begin{equation}
\label{AH3}
\|\Phi\|_{\rm Lip}=\|\Phi\|\le 1.
\end{equation}
Furthermore, by (\ref{AH2}),
\begin{eqnarray}
\label{ZA1}
\Phi\circ J_{pq}\circ \Gamma(f)
&=&\sum_{i=0}^{N-1}\g\circ\b\circ f(i)\,\Phi\psi_i\nonumber\\
&=&
\sigma_1 \sum_{i=0}^{N-1}\g\circ\b\circ f(i)\,e_i\nonumber\\
&=&\sigma_1 J_{pq}^N(\g\circ\b\circ f).
\end{eqnarray}

Define 
$\eta:D\to \Z[0,N)$ by 
$$
\eta(s)=\min\{i\,|\,s\in D_{ki}\},
$$
and 
$$
\rho:D\times\Z[0,2^{m^*})\to\R \quad \mbox{by}\quad \rho(s,z)=\g(z)\psi_{\eta(s)}(s).
$$
Then
\begin{eqnarray*}
\Gamma(f)(s)&=&\sum_{i=0}^{N-1}\g\circ\b\circ f(i)\,\psi_i(s)\\
&=&\g\circ\b\circ f(\eta(s))\,\psi_{\eta(s)}(s)\\
&=&\rho(s,\b\circ f\circ\eta(s)).
\end{eqnarray*}
So $\Gamma$ is of the needed form (see relation (4) of \cite{Hei03a}, 
with $\kappa=1$) and, by (\ref{EF1}), maps 
$$
\mathcal{B}(L_p^N)\quad
\mbox{into}\quad  \sigma_2 2^{rk}\left(1+2^{-m^*/2}\right)\mathcal{B}(W_p^r(D)).
$$
By (\ref{AH3}), Lemma 2 and Corollary 1 of \cite{Hei03a}, and Lemma 6(iii) of 
\cite{Hei01},
\begin{eqnarray}
\lefteqn{e_{2n}^\q(\Phi\circ J_{pq}\circ\Gamma,\mathcal{B}(L_p^N))}&&\nonumber\\
&\le& e_{2n}^\q(J_{pq}\circ\Gamma,\mathcal{B}(L_p^N))\nonumber\\
&\le& e_n^\q\left(J_{pq},\sigma_2 2^{rk}\left(1+2^{-m^*/2}\right)
\mathcal{B}(W_p^r(D))\right)\nonumber\\
&=&\sigma_2 2^{rk}\left(1+2^{-m^*/2}\right)e_n^\q(J_{pq},\mathcal{B}(W_p^r(D))).
\label{AH1}
\end{eqnarray}
Using (\ref{N9}) again, we infer
$$
\sup_{f\in\mathcal{B}(L_p^N)} \|J_{pq}^Nf-J_{pq}^N(\g\circ\b\circ f)\|_{L_q^N}
=\|f-\g\circ\b\circ f\|_{L_q^N}\le 2^{-m^*/2},
$$
and hence, by   Lemma 6(i) and (ii) of \cite{Hei01},
(\ref{ZA1}), and (\ref{AH1}),
\begin{eqnarray*} 
e_{2n}^\q(J_{pq}^N,\mathcal{B}(L_p^N))
&\le&  e_{2n}^\q(J_{pq}^N\circ\overline{\g\circ\b},\mathcal{B}(L_p^N))+2^{-m^*/2}\\
&=& \sigma_1^{-1}e_{2n}^\q(\Phi\circ J_{pq}\circ\Gamma,\mathcal{B}(L_p^N))+2^{-m^*/2}\\
&\le& \sigma_1^{-1}\sigma_2 2^{rk}\left(1+2^{-m^*/2}\right)
e_n^\q(J_{pq},\mathcal{B}(W_p^r(D)))+2^{-m^*/2}\\
&\le& c N^{r/d} e_n^\q(J_{pq},\mathcal{B}(W_p^r(D)))+2^{-m^*/2},
\end{eqnarray*}
where $\overline{\g\circ\b}$ stands for 
$$
(\g\circ\b,\dots,\g\circ\b):\R^N\to \R^N.
$$
Since $m^*$ can be made arbitrarily large, we get
\begin{equation}
\label{XB1}
e_{2n}^\q(J_{pq}^N,\mathcal{B}(L_p^N))\le c N^{r/d} e_n^\q(J_{pq},\mathcal{B}(W_p^r(D))).
\end{equation}
For the case $r/d>2/p-2/q$,  we choose
$k=\lceil d^{-1}(\log(n/c_0)+1)\rceil$, where $c_0$ is the constant from 
Proposition 6 of \cite{Hei03a}, which can be assumed to satisfy $0<c_0\le 1$.
 It follows that 
\begin{equation}
\label{N6}
c_0 2^{-d}N = c_0 2^{-d}2^{dk}\le 2n\le c_0 2^{dk} = c_0 N.
\end{equation}
Now the lower bounds (\ref{XA1a}), (\ref{XA2a}), and  (\ref{XA3a}) follow from 
(\ref{XB1}), (\ref{N6}), and Proposition 6 of \cite{Hei03a}. 
In the case $r/d\le 2/p-2/q$, which implies $p<q$, we set 
$k=\lceil d^{-1}(\log(n^2/c_0)+1)\rceil$.
Consequently,
\begin{equation}
\label{N6a}
c_0 2^{-d}N = c_0 2^{-d}2^{dk}\le 2n^2\le c_0 2^{dk} = c_0 N.
\end{equation}
Relation (\ref{XA4a}) results 
from (\ref{XB1}), (\ref{N6a}), and Proposition 4 in \cite{Hei03a}.
\end{proof}

\section{Comments}
The algorithm  we presented was
optimal with respect to the number of queries. (Although parts of the algorithm 
occur only
in an implicit way, through the use of properties of $e_n^\q$ numbers, it is 
straightforward to transform all upper bound proofs into algorithmic details.)
Let us now consider its cost in the bit model of computation. 
Here we assume that $n$ and $N$ are powers of 2. We use the respective remarks 
about bit cost made in Section 5 of \cite{Hei03a}. 

For $p\ge q$ classical approximation suffices. For $p<q$ the problem 
 splits into the classical computation of $P_{l_0}$ and the approximation
of $J_{pq}^{N_l}$ for $l=l_0,\dots,l^*-1=2l_0-1=\mathcal{O}(\log n)$
using $n_l$ queries (see the proof Proposition \ref{pro:8} for these numbers).
To increase the respective success probabilities appropriately, 
we have to repeat these
approximations $\nu_l$ times on level $l$, and we have 
$\nu_l=\mathcal{O}(\log \log n)$. The total number of 
queries is $n$ (or $\wt{n}=\mathcal{O}(n)$ if considered before scaling), 
that of quantum gates is
$$
\mathcal{O}\left(\sum_{l=l_0}^{l^*-1}\nu_l n_l\log N_l\right)=\mathcal{O}(n\log n).
$$
The algorithm needs $\mathcal{O}(\log n)$ qubits and
$$
\mathcal{O}\left(\sum_{l=l_0}^{l^*-1}
\nu_l n_l^2 N_l^{-1}\max(\log(n_l/\sqrt{N_l}),1)^{-1}
\right)=\mathcal{O}(n/\log n)
$$
measurements.
To compute $P_{l_0}f$ classically, we need $\mathcal{O}(n)$ function values
and $\mathcal{O}(n\log n)$ classical bit operations.
For the approximations on the levels a total of  
$$
\mathcal{O}\left(\sum_{l=l_0}^{l^*-1}
\nu_l n_l^2 N_l^{-1}\log N_l
\right)=\mathcal{O}(n\log n).
$$
classical bit operations is required.
This does not yet take into account
the classical computation of the vector analogue of the median. Let
us assume that we apply the constructive procedure described
after Corollary 1 of \cite{Hei03a}. At level $l$ we have to compute the 
norm of $\nu_l^2$ vectors in $L_q^{N_l}$ with at most
$$
\mathcal{O}\left(
  n_l^2 N_l^{-1}\max(\log(n_l/\sqrt{N_l}),1)^{-2}\right)
$$
non-zero coordinates. This amounts to
$$
\mathcal{O}\left(\log n\sum_{l=l_0}^{l^*-1}
 \nu_l^2 n_l^2 N_l^{-1}\max(\log(n_l/\sqrt{N_l}),1)^{-2}
\right)=\mathcal{O}(n\log\log n)
$$
classical bit operations.
We see that the overall quantum bit cost differs by at most a logarithmic
factor from the quantum query cost $\Theta(n)$.

The concrete form of the output of the algorithm depends on the structure of $P$.
If $P$ is, for example, tensor product Lagrange interpolation, then the output
is a sum of $\mathcal{O}(\log n)$ piecewise polynomial functions, with 
$\mathcal{O}(n)$ pieces for the classical part $P_{l_0}f$ and 
$$
\mathcal{O}\left(
  n_l^2 N_l^{-1}\max(\log(n_l/\sqrt{N_l}),1)^{-2}
\right)
$$
pieces not identical to zero on level $l$, that is, a total of 
$$
\mathcal{O}\left(n + \sum_{l=l_0}^{l^*-1}
 n_l^2 N_l^{-1}\max(\log(n_l/\sqrt{N_l}),1)^{-2}
\right)=\mathcal{O}(n)
$$
nontrivial pieces, with each point of $D$ being contained in at most 
$\mathcal{O}(\log n)$ pieces.

We summarize the results on the approximation of 
$J_{pq}:\mathcal{B}(W_p^r([0,1]^d))\to L_q([0,1]^d)$
in a table and compare them with
the respective known quantities in the classical deterministic and randomized settings
(see \cite{Hei93} and the bibliography therein).  Recall 
that we always assume $r/d>1/p$. The respective entries of the table 
give the minimal errors, constants and logarithmic factors are suppressed.

\[
\begin{array}{l|l|l|l}
\qquad J_{pq}
& \ \mbox{deterministic}\  & \, \mbox{random}\, & \, \mbox{quantum}\,\\ \hline 
&&&\\
\;1\le p < q\le \infty,&\, n^{-r/d+1/p-1/q} & \, n^{-r/d +1/p-1/q}  
& \, n^{-r/d}\\
 r/d\ge 2/p-2/q&&&\\
&&&\\
\;1\le p < q\le \infty,&\, n^{-r/d+1/p-1/q} & \, n^{-r/d +1/p-1/q}  
& \, n^{-2r/d+2/p-2/q}\\
 r/d< 2/p-2/q&&&\\
 &&&\\
\;1\le q \le p \le \infty&\, n^{-r/d} & \, n^{-r/d}  
& \, n^{-r/d}\\
\end{array}
\]

We observe a possible improvement of $n^{-1}$ 
(for $p=1$, $q=\infty$) over the classical deterministic and randomized 
case (which is essentially 
a squaring of the classical rate for $r/d$ close to 1).
This is the maximal speedup over the randomized case 
observed so far in natural numerical problems (the same speedup was first found
in \cite{Hei01b} for integration 
of functions from $W_1^r(D)$). We also see that there are regions of
 the parameter domain where
the speedup is smaller, and others, where there is no speedup at all.


\begin{thebibliography}{BHMT00}


\bibitem{Ada75}
R.\ A.\ Adams, 
Sobolev Spaces,
Academic Press, New York, 1975. 


\bibitem{Cia78}
P.~G. Ciarlet, The Finite Element Method for Elliptic Problems,
North-Holland, Amsterdam, 1978.


\bibitem{Hei93}
S. Heinrich,
Random approximation in numerical analysis,
in: K.~D. Bierstedt, A.~Pietsch, W.~M. Ruess, D.~Vogt (Eds.),
 Functional Analysis, Marcel {D}ekker, New York, 1993, 123 -- 171.



\bibitem{Hei01}
S. Heinrich, 
Quantum summation with an application to integration, 
Journal of Complexity 18 (2002), 1--50, see also 
http://arXiv.org/abs/quant-ph/0105116.


\bibitem{Hei01b} 
S.\ Heinrich, 
Quantum integration in 
Sobolev classes, J. Complexity 19 (2003), 19--42, see also 
http://arXiv.org/abs/quant-ph/0112153.
 


\bibitem{Hei03a}
S.\ Heinrich, 
Quantum Approximation I. Embeddings of Finite Dimensional $L_p$ Spaces, 2003.


\bibitem{HN01b} 
S.\ Heinrich, E.\ Novak, On a problem in quantum summation,
J. Complexity  19 (2003), 1--18, see also http://arXiv.org/abs/quant-ph/0109038.


\bibitem{Mai75} 
V. E. Maiorov, Discretization of the problem of diameters,
Usp. Mat. Nauk 30, No. 6 (186) (1975), 179--180. 



\bibitem{NSW02}
E. Novak, I. H. Sloan, H. Wo\'zniakowski,
Tractability of approximation for weighted Korobov spaces on 
classical and quantum computers, 2002, see  http://arXiv.org/abs/quant-ph/0206023.



\bibitem{Tri95}
H.\ Triebel,
Interpolation Theory, Function Spaces, Differential Operators, 2nd ed., 
Barth, Leipzig, 1995.


\end{thebibliography}
\end{document}